\documentclass[a4paper]{article}

\usepackage{INTERSPEECH2022}
\usepackage{subfigure}
\usepackage{url}

\usepackage[hidelinks]{hyperref}
\usepackage{color}

\newcommand{\drawpdf}[4]{ 
 \begin{figure}[#1]
 \centering
 \includegraphics[width=#2]{#3.pdf}
 \vspace{-2pt}
 \caption{#4}
 \label{fig:#3}
 \end{figure}
}

\newcommand{\Argmax}{\mathop {{\rm argmax}}}

\title{
Human-in-the-loop Speaker Adaptation for DNN-based Multi-speaker TTS
}
\name{Kenta Udagawa, Yuki Saito, Hiroshi Saruwatari}
\address{
 Graduate School of Information Science and Technology, The University of Tokyo, Japan.
 }
\email{udagawa-kenta338@g.ecc.u-tokyo.ac.jp, yuuki\_saito@ipc.i.u-tokyo.ac.jp}

\begin{document}

\setlength{\abovedisplayskip}{3pt} 
\setlength{\belowdisplayskip}{3pt} 
\setlength\floatsep{5pt} 
\setlength\intextsep{3pt} 
\setlength\textfloatsep{3pt} 
\setlength\abovecaptionskip{5pt} 
\setlength{\dbltextfloatsep}{6pt} 
\setlength{\dblfloatsep}{2pt}

\maketitle

\begin{abstract}
This paper proposes a human-in-the-loop speaker-adaptation method for multi-speaker text-to-speech. With a conventional speaker-adaptation method, a target speaker's embedding vector is extracted from his/her reference speech using a speaker encoder trained on a speaker-discriminative task. However, this method cannot obtain an embedding vector for the target speaker when the reference speech is unavailable. Our method is based on a human-in-the-loop optimization framework, which incorporates a user to explore the speaker-embedding space to find the target speaker's embedding. The proposed method uses a sequential line search algorithm that repeatedly asks a user to select a point on a line segment in the embedding space. To efficiently choose the best speech sample from multiple stimuli, we also developed a system in which a user can switch between multiple speakers' voices for each phoneme while looping an utterance. Experimental results indicate that the proposed method can achieve comparable performance to the conventional one in objective and subjective evaluations even if reference speech is not used as the input of a speaker encoder directly.
\end{abstract}

\vspace{-5pt}
\section{Introduction}
\vspace{-3pt}
Text-to-speech (TTS) based on deep neural networks (DNNs)~\cite{zen13dnn,shen18} has attracted attention as a technique for synthesizing a high-quality speech from text. Multi-speaker TTS, such as Deep Voice 2~\cite{arik17deepvoice2}, aims to synthesize various speaker's voices using a single TTS model. A common method for multi-speaker TTS is to use a lookup table of speaker embeddings, i.e., a trainable matrix representing each speaker's real-valued ID. Because most parameters of the TTS model with this method are shared among different speakers, one can build a multi-speaker TTS system using a small amount of data per speaker. However, with such multi-speaker TTS methods using a speaker-embedding table, is difficult to synthesize the voice of an unseen speaker who is not included in the training data. Speaker adaptation for multi-speaker TTS~\cite{wu15dnnadapt, taigman2018voiceloop, arik18} is a technique for synthesizing the speech of such unseen speakers.

Speaker adaptation for multi-speaker TTS can be interpreted as the task of searching the speaker-embedding space for a vector that can reproduce the target speaker's voice. With a transfer-learning-based speaker-adaptation method~\cite{jia2018transfer}, a few seconds of the target speaker's reference speech is prepared, and the target speaker's embedding vector is estimated by feeding the reference speech to the speaker encoder pre-trained on the speaker-verification task. This method can synthesize high-quality speech for the target speaker from only a small amount of reference speech while reducing the cost of collecting high-quality speech and transcription pair data for TTS model training. However, this method cannot be applied when reference speech from the target speaker is unavailable.

We propose a speaker-adaptation method that incorporates human speech perception to explore the target speaker's embedding. The proposed method is based on a sequential line search (SLS) algorithm~\cite{koyama2017sls}, in which a user repeatedly selects a point on a presented line segment in the search parameter space by manipulating a single slider. In the speaker-adaptation setting, the user selects the voice closest to the target speaker from voices synthesized from multiple speaker embeddings on a line segment of the speaker-embedding space. The proposed method synthesizes multiple voices in advance from multiple speaker embeddings to improve the exploration efficiency, and the user selects one voice using a system with which the user can switch the played-back voice for each phoneme while looping an utterance. The proposed method can be used without the target speaker's reference speech because it feeds back only human-preference data for searching the target speaker's embedding. Experimental results indicate that the proposed method can synthesize speech with the same or better quality than the conventional method, depending on the user and test speaker.

\vspace{-5pt}
\section{Conventional speaker adaptation}
\vspace{-3pt}
\subsection{Multi-speaker text-to-speech synthesis}
\vspace{-3pt}
\label{sec:multi}
We first give an overview of the multi-speaker TTS system used in this paper. As shown in Figure~\ref{fig:multi-tts}, this system consists of two DNNs: a synthesizer that generates a mel spectrogram from a phoneme sequence and speaker-embedding vector and a vocoder that synthesizes a speech waveform from the generated mel spectrogram. A synthesizer is trained on pairs of text and audio, and a vocoder is trained separately on pairs of mel spectrograms and speech waveforms. The structure of this TTS system is common to both conventional and proposed methods.

\drawpdf{t}{\linewidth}{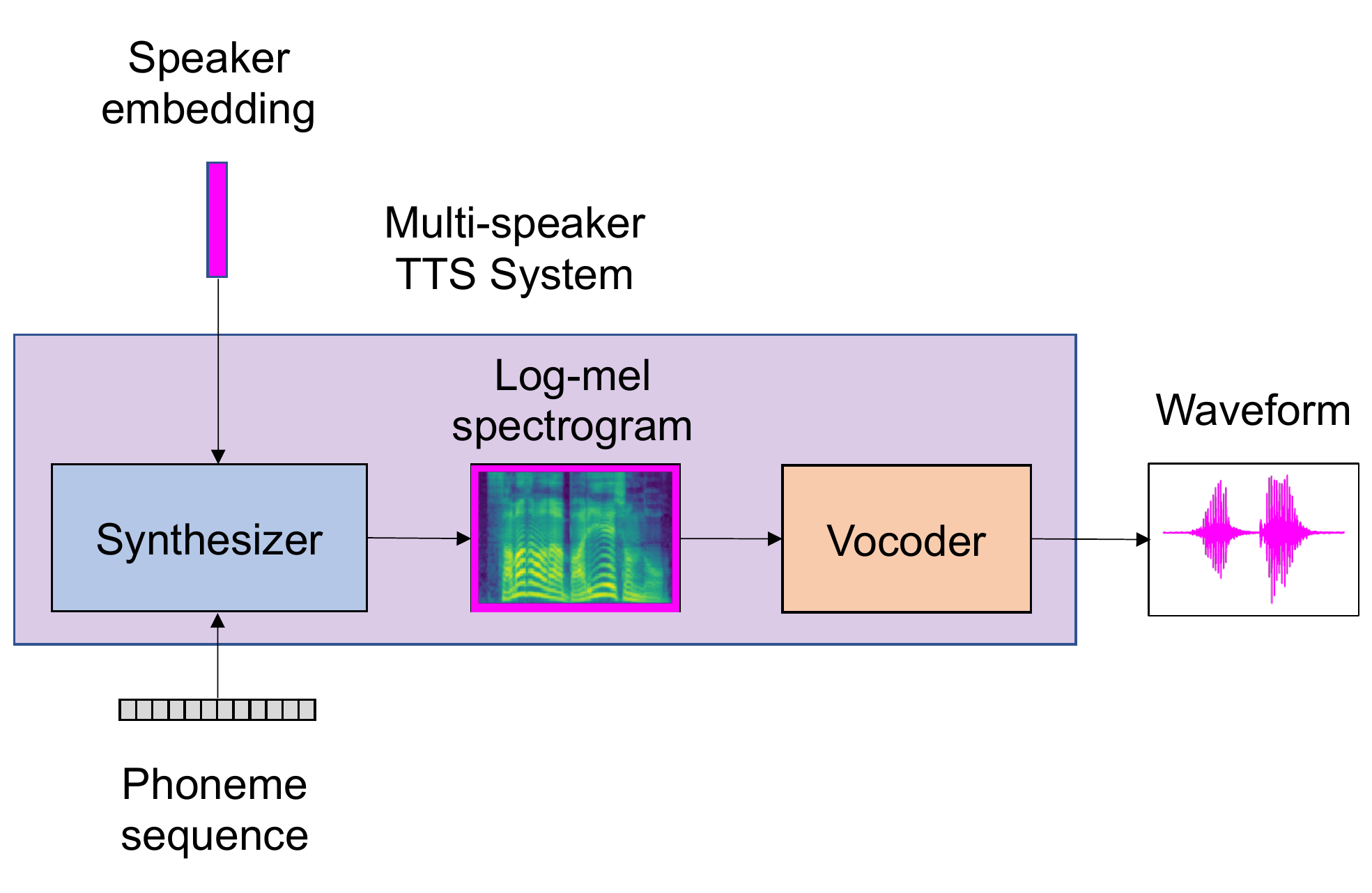}
{Overview of multi-speaker TTS system}

\vspace{-5pt}
\subsection{Speaker adaptation based on transfer learning from speaker verification}
\vspace{-3pt}
With the speaker-adaptation method based on transfer learning from speaker verification~\cite{jia2018transfer}, the target speaker's embedding is obtained using a speaker encoder that extracts the speaker-embedding vector from the speech waveform. The speaker encoder is trained to optimize the generalized end-to-end (GE2E) loss for speaker verification~\cite{wan2018generalized} separately from the synthesizer and vocoder networks. The training using GE2E loss makes speaker embeddings extracted from the same speaker's speech waveforms closer together and moves different speakers' embeddings farther apart. The synthesizer is trained to predict a mel spectrogram from given text with the speaker embedding extracted with the trained speaker encoder. During inference, the target speaker's embedding is extracted from a few seconds of the target speaker's reference speech, then the extracted speaker embedding and text are fed into the trained multi-speaker TTS system to synthesize a speech waveform with the voice of the target speaker.

\vspace{-5pt}
\section{Human-in-the-loop speaker adaptation}
\vspace{-3pt}
\subsection{Sequential line search}
\label{sls_exp}
We first describe the SLS algorithm~\cite{koyama2017sls}, which is the search algorithm used with the proposed method.
Let $D$ be the dimensionality of the search parameters, $\mathcal{X}=[0, 1]^D$ is search parameter space, and $g:\mathcal{X}\rightarrow \mathbb{R}$ is user's perceptual preference function, respectively.
The SLS algorithm repeats the process of having the user select a parameter point from a line segment in $\mathcal{X}$ to obtain ${\bf x}^{*}$, which is defined as:
\begin{align}
 {\bf x}^{*} = \Argmax_{{\bf x} \in \mathcal{X}} g({\bf x}).
\end{align}
Let $S_t$ be the line segment proposed for the $t$th step in the SLS algorithm, ${\bf x}^{+}_{t-1}$, and ${\bf x}^{\textrm{EI}}_{t-1}$ be the endpoints of $S_t$, respectively.
In the original SLS setting, the initial endpoints ${\bf x}^{+}_{0}, {\bf x}^{\textrm{EI}}_{0}$ are two random points in $\mathcal{X}$. The endpoints ${\bf x}^{+}_{t}, {\bf x}^{\textrm{EI}}_{t}$ of the line segment $S_{t+1}$ are determined from the following equations:
\begin{align}
\label{x1_def}
{\bf x}^{+}_{t} &= {\bf x}^{\textrm{chosen}}_t, \\
\label{x2_def}
{\bf x}^{\textrm{EI}}_{t} &= \Argmax_{{\bf x} \in \mathcal{X}} a^{\textrm{EI}}({\bf x}; D_t).
\end{align}
Here, ${\bf x}^{\textrm{chosen}}_t$ represents the point selected by the user from $S_t$, and this definition of ${\bf x}^{+}_{t}$ was proposed in  ~\cite{koyama2020sg}, $D_t$ represents the user's selection up to the $t$th step, and $a^{\textrm{EI}}(\cdot)$ is an acquisition function based on expected improvement (EI)~\cite{jones1998efficient} and defined as:
\begin{align}
 a^{\textrm{EI}}({\bf x}; D_t) = \mathbb{E}[\max\{ g({\bf x}) - g^{+} , 0\}],
\end{align}
where $g^{+}$ is the best preference function among the observation points. To calculate $a^{\textrm{EI}}({\bf x}; D_t)$, a Gaussian process prior is assumed on $g(\cdot)$ to make the preference-function value $g({\bf x}_{*})$ at the unobserved point ${\bf x}_{*}$ follow the normal distribution:
\begin{align}
 g({\bf x}_{*}) \sim \mathcal{N}(\mu({\bf x}_{*}), \sigma({\bf x}_{*})).
\end{align}
The mean $\mu_t({\bf x}_{*})$ and variance $\sigma_t({\bf x}_{*})$ of the distribution depend on the preference-function value ${\bf g} = (g({\bf x}_1), \cdots, g({\bf x}_{N_t}))^{\top }$ at observation point $\{{\bf x}_i\}_{i=1}^{N_t}$ and hyperparameters ${\boldsymbol \theta}$. The preference-function value at the observation point ${\bf g}$ and hyperparameters ${\boldsymbol\theta}$ are estimated by maximum a posteriori estimation. This allows the probability distribution of the preference-function value $g({\bf x}_{*})$ at the unobserved point ${\bf x}_{*}$ to be expressed in closed form and the acquisition function $a^{\bf \textrm{EI}}({\bf x}; D_t)$ to be calculated. Therefore, the endpoints ${\bf x}^{+}_{t}, {\bf x}^{\textrm{EI}}_{t}$ of line segment $S_{t+1}$ can be determined from Equations (\ref{x1_def}) and (\ref{x2_def}).

\drawpdf{t}{\linewidth}{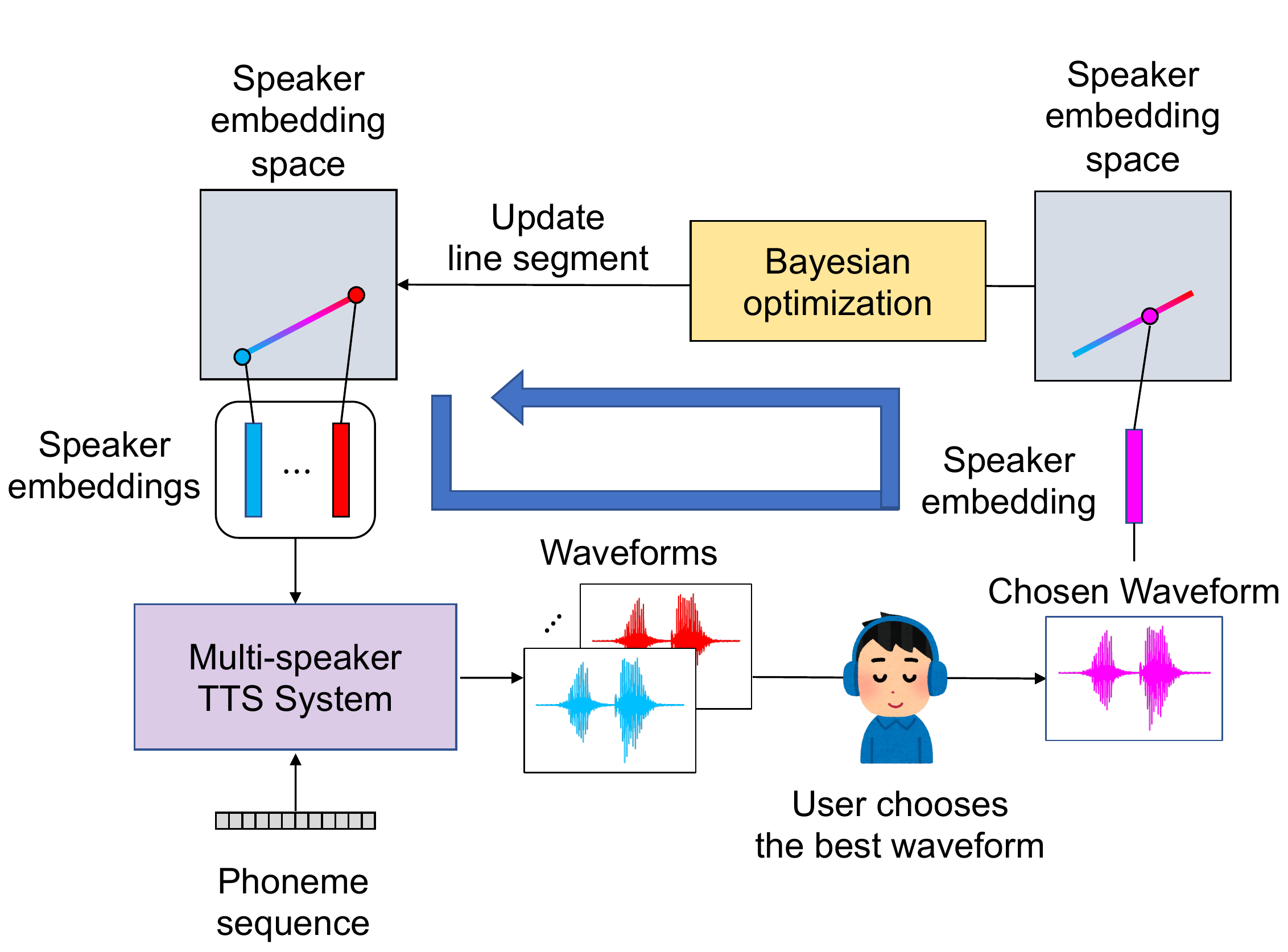}
{Overview of proposed speaker-adaptation method}

\vspace{-5pt}
\subsection{SLS-based speaker adaptation}
\vspace{-3pt}
Our speaker-adaptation method uses the SLS algorithm to search for the target speaker's embedding. An overview of the proposed method is shown in Figure \ref{fig:prop+allow}. The user selects the waveform closest to the target speaker's voice from multiple synthesized speech waveforms. When the user perceives non-time-series data such as images, a single piece of data can be perceived instantly. However, for time-series data such as voice, it takes several seconds to perceive an entire piece of data. For the user to efficiently select the best waveform from multiple waveforms, we developed a system in which the user can switch the played-back voice for each phoneme while looping an utterance. With this system, the user selects the preferred speech waveform from a continuously changing voice by manipulating a single slider.

Since the proposed human-in-the-loop speaker-adaptation method does not use the reference voice as input for the speaker encoder, it has the advantage of synthesizing the voice of the speaker imagined by the user. This means that the method can be used even when reference voices are difficult to obtain, such as for deceased or inarticulate speakers. The proposed method also has possible applications such as cross-lingual speaker adaptation~\cite{xin20_interspeech} and audiobook speech synthesis~\cite{nakata21_ssw}.

\vspace{-5pt}
\subsection{Initial line segment and search-parameter space}
\vspace{-3pt}
To search for high-quality synthetic speech\footnote{In our preliminary experiments, we observed that the quality of the synthesized speech is lower when the speaker embedding is far from the distribution of the training data.}, the endpoints of the initial line segment with the proposed method are set to the mean male-speaker embedding and mean female-speaker embedding. In addition, the search-parameter space $\mathcal{X}$ is set to the quantile $\mathcal{Q}$ of the training data.
Let $q_i$ be the $i$th component of the SLS search parameter ${\bf q}\in \mathcal{Q}=[0, 1]^{16}$, $e_i$ be $i$th component of the speaker embedding ${\bf e}\in [0, 1]^{16}$, and $e ^{\mathrm{train}}_{i,j}$ be $j$th smallest $i$-component of the $N$ speaker embeddings in the training data for the synthesizer ($e^{\mathrm{train}}_{i,0}=-\infty, e^{\mathrm{train}}_{i,N+1}=\infty$), respectively.
The transformation $\mathrm{Conv}$ from $e_i$ to $q_i$ and its inverse function $\mathrm{IConv}$ $q_i$ to $e_i$ are defined as follows (where $[\Vec{\cdot}]$ is a floor function):

\begin{eqnarray}
 \mathrm{Conv}(e_i) &=& j/N, \ \ (e^{\mathrm{train}}_{i,j} \leq e_i < e^{\mathrm{train}}_{i,j+1} ), \\
 \mathrm{IConv}(q_i) &=& e^{\mathrm{train}}_{i,[q_i (N-1) + 1]}.
\end{eqnarray}

\vspace{-5pt}
\section{Experiments}
\vspace{-3pt}
\subsection{Experimental conditions}
\vspace{-3pt}
For training the speaker encoder, we used the Japanese speech data included in the Corpus of Spontaneous Japanese (CSJ)~\cite{maekawa2003corpus}. The CSJ contains 660 hours of speech data from 1417 Japanese speakers (947 men and 470 women). The CSJ audio data were resampled to 16~kHz, and the frame shift was set to 10~ms. For training, validating, and testing the synthesizer, we used the Japanese speech data included in the parallel data of the Japanese Versatile Speech (JVS) corpus~\cite{takamichi2020E1950}. The parallel data of the JVS corpus contain 22 hours of speech data from 100 Japanese speakers (49 men and 51 women; 100 sentences per speaker). The validation and test data were 100 utterances of 10 speakers (5 men and 5 women) randomly sampled from the parallel data of the JVS corpus, and the training data were 20 hours of speech data from the remaining 90 speakers. As corner cases in speaker-adaptation difficulty, we selected four speakers for the test data on the basis of the speaker's average subjective similarity~\cite{saito21taslp} to speakers in the training data: the highest male (``jvs078'') and female (``jvs060'') and the lowest male (''jvs005'') and female (``jvs010''). We excluded data containing annotation errors and recording failure\footnote{ The preprocessing of the JVS corpus was on the basis of the repository at \url{https://github.com/Hiroshiba/jvs_hiho}}. The JVS speech data were resampled to 22.05~kHz to match the settings of neural vocoding, and the frame shift was set to 12~ms.

For the synthesizer, we used the open source implementation of FastSpeech 2~\cite{ren2021fastspeech} published by ming024\footnote{\url{https://github.com/ming024/FastSpeech2}}. FastSpeech 2 predicted the 80-dimensional mel spectrogram from Japanese phonemes with the aid of the variance adaptors that predicted the F0 and energy of synthetic speech. The F0 was estimated using the WORLD vocoder~\cite{morise16world,morise16d4c}. In the FastSpeech 2 training, we used the Adam optimizer~\cite{kingma14adam} with 4,000 Warmup~\cite{goyal2017accurate} steps, an initial learning rate of 0.0625, batch size of 8, and 50,000 training steps.

We used the same architecture of a speaker encoder as with the conventional method~\cite{jia2018transfer}. However, we reduced the speaker embedding's dimensionality from 256 to 16 for the proposed method to make the Bayesian-optimization-based SLS algorithm work well~\cite{moriconi2020high}. We also changed the speaker-embedding layer's activation function from a rectified linear unit (ReLU)~\cite{glorot11relu} with L2-normalization to sigmoid. This is because with the SLS algorithm a hypercube of $[0, 1]^D$ as the $D$-dimensional search parameter space is assumed\footnote{In our preliminary experiments, we observed that these modifications did not affect the quality of synthetic speech with the conventional speaker adaptation method.}. We denote {\bf EMB256dim } and {\bf EMB16dim} as the speaker embeddings used for the conventional and proposed methods, respectively. Both types of speaker embedding were projected to the 256-dimensional vector space through the two fully connected layers: 256 hidden units with the ReLU activation function and 256 output units with the linear activation function, then added to the output of text encoder in the synthesizer. In the speaker-encoder training, we used the Adam optimizer with a learning rate of 0.0001, batch size of 8, and 1,000,000 training step. The vocoder for the speech-waveform synthesis was the generator\_universal model of HiFi-GAN~\cite{kong2020hifigan} included in the FastSpeech 2 repository published by ming024.

We used the SLS setting in which the line segment is extended by 1.25 times before presenting it to the user. The number of points that can be selected from the SLS line segment was 20. The hyperparameters ${\boldsymbol \theta}$ were set to default, i.e., same as the repository at \url{https://github.com/yuki-koyama/sequential-line-search}.

The mean absolute error (MAE) between the natural and synthetic mel spectrograms was used as an objective evaluation metric regarding the quality of the speech by speaker adaptation. The mel spectrogram MAE was calculated for speech synthesized by giving FastSpeech 2 the phoneme duration of the natural speech. With both the proposed and conventional methods, noise was observed in the pause interval of the speech synthesized by FastSpeech 2. Therefore, the paused part of the mel spectrogram was masked when synthesizing speech and omitted when calculating the mel spectrogram MAE.

\vspace{-5pt}
\subsection{Human-in-the-loop adaptation experiment}
\vspace{-3pt}
\label{human_subeval}
We conducted an experiment in which participants used the proposed method to search for the target speakers' voices. The participants were instructed to select the synthesized voice they thought was closest to the target speaker's voice at each step. The number of SLS steps was set to 30 so that the experimental time per speaker adaptation would be 15--25 minutes. The participants were eight members of a laboratory specializing in speech and acoustic processing. Although the proposed method can execute speaker adaptation to the target speaker in the user's mind without the reference voice, in this experiment, to make the participants fully aware of the target speaker's voice, we allowed them to listen to the reference voice when using the proposed method. However, we prohibited the participants from playing the synthesized voice and reference voice simultaneously and comparing them to simulate a situation in which the target speaker's voice is in the user's mind. We used the VOICEACTRESS100\_001 utterance in the parallel data of the JVS corpus for the speaker-embedding search with the proposed method. The duration of the speech VOICEACTRESS100\_001 in the test speaker was about eight seconds. The following two methods were compared with the proposed method.

\begin{enumerate}
 \item
 {\bf TL}: The mean of ``EMB256dim'' speaker embeddings extracted from all utterances per test speaker. This represents the conventional method~\cite{jia2018transfer}.
 \item
 {\bf Mean-Speaker}: The mean of ``EMB16dim'' speaker embedding extracted from all the JVS training data of the same sex as the target speaker. This represents the initial value of the SLS algorithm with the proposed method. The difference between the proposed method and this means a change in a quality of synthesized voice due to human manipulation.
\end{enumerate}

\drawpdf{t}{0.98\linewidth}{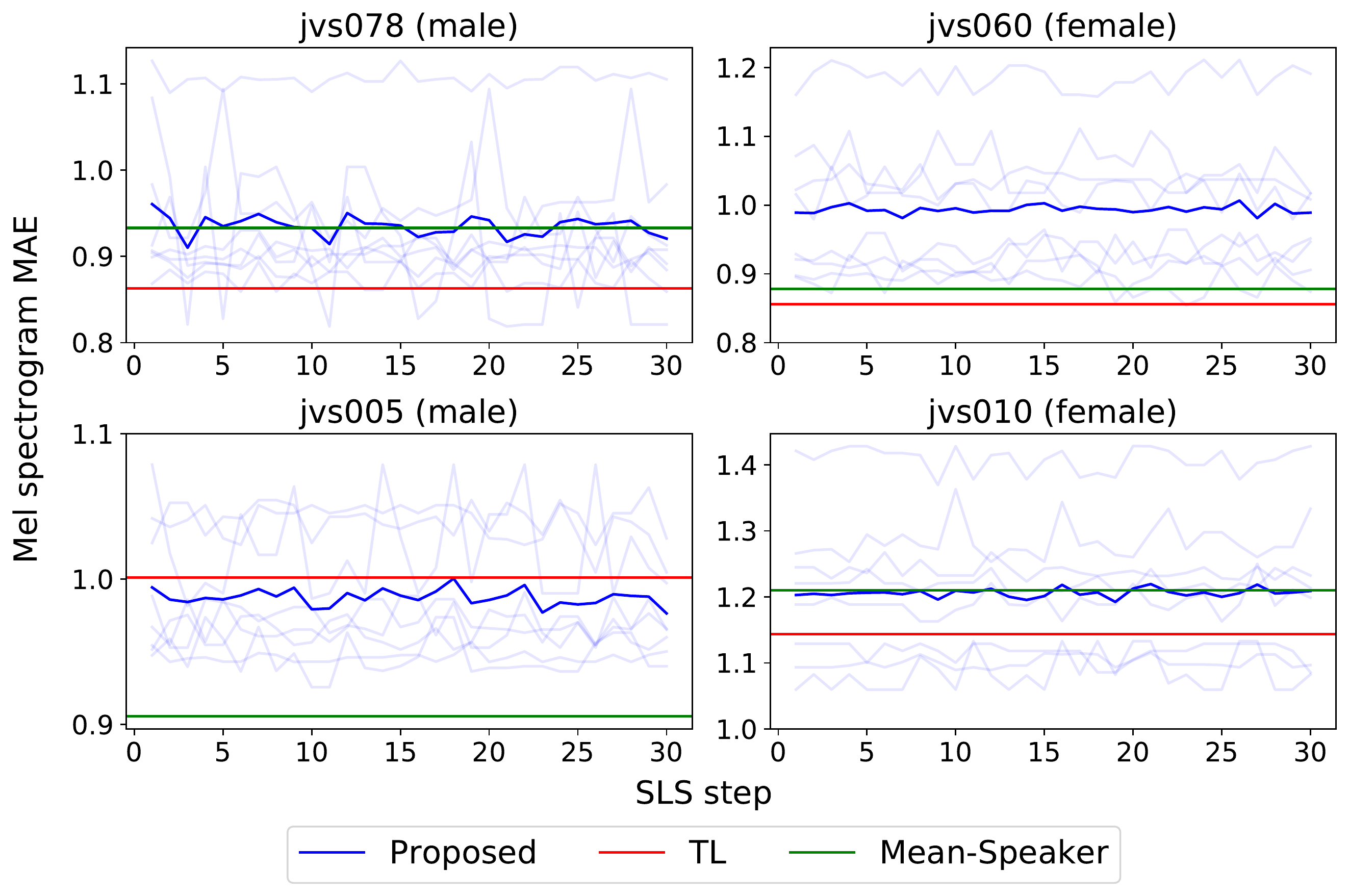}
{Mel spectrogram MAE per step when eight participants used the proposed method. }

To examine the quality of the voice selected by a participant at each step, we calculated the mel spectrogram MAE of the speech for embedding exploration per step. Figure~\ref{fig:sls2_human} shows the mel spectrogram MAE of the speech for embedding exploration per step when the proposed method was used up to 30 steps by the eight participants. The light blue line in Figure~\ref{fig:sls2_human} represents the results from one participant, and the dark blue line represents the average results from all eight participants. The red and green lines represent the mel spectrogram MAE of ``TL'' and ``Mean-Speaker,'' respectively. The light blue line with the lowest mel spectrogram MAE shows that the proposed method can synthesize speech comparable to the conventional method in terms of objective evaluation. However, if we focus on the dark blue line (the average results from the eight participants), we can see that the mel spectrogram MAE does not tend to improve by human manipulation for all test speakers. The possible reasons for this are as follows. (1) The manipulation task was difficult because the search parameters (i.e., speaker embedding) were too abstract. (2) The mel spectrogram MAE with ``Mean-Speaker'' (i.e., the initial value with the proposed method) was sufficiently low.

\vspace{-5pt}
\subsection{Objective evaluation}
\vspace{-3pt}
\label{sec:mulutt_obj}
As the objective evaluation of the conventional and proposed methods, we computed the mel spectrogram MAE between the natural and synthetic speech by using each test speaker's 100 utterances.
In addition to ``TL’' and ``Mean-Speaker’‘, we defined {\bf SLS-\{Best, Mean, Worst\}} as the speaker embedding whose the mel spectrogram MAE with the utterance for searching was \{minimum, mean, maximum\} among speaker embeddings obtained after the eight participants manipulated the proposed method up to 30 steps.
Table \ref{table:1} shows the results\footnote{The audio samples are available on \href{http://sython.org/demo/udagawa22interspeech/demo_slstts.html}{\color{blue}
this page\color{black}
}.}. The quality of ``SLS-Best'' for speakers other than ``jvs060'' and ``SLS-Mean'' for speakers of ``jvs078'' and ``jvs005'' was comparable to ``TL''. However, for all test speakers, ``SLS-Worst'' was significantly inferior to ``TL'' and ``Mean-Speaker''.
The reason for this could be that the manipulation task was difficult, as discussed in Section \ref{human_subeval}.

\begin{table}[tb]
\footnotesize
\caption{ Mel spectrogram MAE of 100 test utterances. \textbf{Bold} values are best among 5 methods per test speaker. }
\vspace{-15pt}
\label{table:1}
\begin{center}
\begin{tabular}{c|ccc|c|c}
\hline
\hline
& \multicolumn{3}{|c|}{SLS} & & Mean- \\
& Worst & Mean & Best & TL & Speaker \\
\hline
jvs078 & 1.03 & 0.85 & {\bf 0.80} & 0.82 & 0.89 \\
jvs005 & 1.06 & 0.99 & {\bf 0.94} & 0.96 & 0.95 \\
jvs060 & 1.10 & 0.95 & 0.90 & {\bf 0.84} & 0.90 \\
jvs010 & 1.27 & 1.14 & \textbf{1.05} & \textbf{1.05} & 1.14 \\
\hline
\hline
\end{tabular}%
\end{center}
\end{table}

\vspace{-5pt}
\subsection{Subjective evaluation}
\vspace{-3pt}
The naturalness and speaker similarity of 100 utterances of the four test speakers synthesized with the conventional and proposed methods were evaluated on the basis of the five-level mean opinion score (MOS) and degradation MOS (DMOS) tests, respectively. The synthetic voices to be compared were the same as with the five methods compared in Section \ref{sec:mulutt_obj}. In the subjective evaluation of naturalness, natural speech ground-truth (GT) was evaluated in addition to speech synthesized with the five methods, and in the subjective evaluation of speaker similarity, natural speech was used as the reference speech. Subjective evaluation was conducted separately for each of the eight tasks (four test speakers $\times$ \{MOS or DMOS\}). For each task, 50 listeners were recruited by crowdsourcing, and they evaluated the quality of speech samples. The number of speech samples in the evaluation was five utterances (randomly sampled from 100 utterances) $\times$ the number of comparison methods (six for the MOS test and five for the DMOS test). One of the five utterances was used as a dummy voice for the listeners to define the evaluation criteria.

\begin{table}[tb]
\centering
\footnotesize
\caption{ Subjective evaluation results. \underline{Underlined} values are comparable with or better than ``TL'' with statistical significance at 5-percent level. }
\vspace{-5pt}
\label{table:2}
\subtable[MOS of speech naturalness]{
\label{table:2a}
\begin{tabular}{c|ccc|c|c|c}
\hline
\hline
& \multicolumn{3}{|c|}{SLS} & & Mean- & \\
& Worst & Mean & Best & TL & Speaker & GT \\
\hline
jvs078 & 3.29 & 3.46 & 3.48 & 3.67 & \underline{ 3.56} & 4.17 \\
jvs005 & 2.81 & \underline{ 3.63} & \underline{ 3.61 } & 3.56 & \underline{ 3.52 } & 4.12 \\
jvs060 & 3.19 & 3.13 & \underline{ 3.42 } & 3.35 & \underline{ 3.34 } & 4.25 \\
jvs010 & \underline{ 3.48 } & \underline{ 3.62 } & 3.13 & 3.42 & \underline{ 3.58 } & 3.75 \\
\hline
\hline
\end{tabular}%
}
\subtable[DMOS of speaker similarity]{
\label{table:2b}
\begin{tabular}{c|ccc|c|c}
\hline
\hline
& \multicolumn{3}{|c|}{SLS} & & Mean- \\
& Worst & Mean & Best & TL & Speaker \\
\hline
jvs078 & 1.90 & \underline{ 2.89 } & \underline{ 3.04 } & 2.71 & 1.95 \\
jvs005 & 1.47 & 2.12 & \underline{ 2.54 } & 2.33 & \underline{ 2.67 } \\
jvs060 & 2.25 & 2.19 & 2.68 & 3.20 & 2.98 \\
jvs010 & 1.54 & 2.03 & \underline{ 3.08 } & 2.98 & 1.96 \\
\hline
\hline
\end{tabular}%
}
\end{table}

The results of the subjective evaluation of naturalness are shown in Table~\ref{table:2a}. First, ``Mean-Speaker'' was about as natural as ``TL''. This indicates that the proposed method using the SLS algorithm with ``Mean-Speaker’’ as the initial value can achieve the same level of naturalness as the conventional method. Although ``SLS-Worst'' of ``jvs005'' was significantly inferior to the conventional method, the naturalness of the proposed method was comparable to the conventional method in several cases.

The results of the subjective evaluation of speaker similarity are shown in Table~\ref{table:2b}. With the proposed method, ``SLS-Best'' outperformed the conventional method for all speakers except ``jvs060'', and ``SLS-Mean'' was comparable to the conventional method for ``jvs078''. The results suggest that the proposed method can achieve the same quality of speaker adaptation as the conventional method even if reference speech is not used as the input of a speaker encoder. However, there were several cases in which the DMOS of the proposed method was below ``Mean-Speaker''. This means that the speaker similarity was lowered by the human search than the speech contained in the initial line. The reason for this could be that the manipulation task was difficult, as discussed in Section \ref{human_subeval}.

\vspace{-5pt}
\section{Conclusions}
\vspace{-3pt}
We proposed a human-in-the-loop speaker-adaptation method that does not input reference speech to the speaker encoder and uses only human preference to find the target speaker's embedding. The experimental results indicate that the proposed method can synthesize speech with the same quality as the conventional method in both objective and subjective evaluations. To make the user's manipulation of the parameters more intuitive for non-expert users, we will investigate a method for providing explainability to the speaker embedding.

\vspace{-5pt}
\section{Acknowledgements}
\vspace{-3pt}
This work was supported by JST, Moonshot R\&D Grant Number JPMJPS2011.

\bibliographystyle{IEEEtran}

\bibliography{template}

\end{document}